\documentclass[article,shortnames,nojss]{jss}\usepackage[]{graphicx}\usepackage[]{xcolor}
\makeatletter
\def\maxwidth{ %
  \ifdim\Gin@nat@width>\linewidth
    \linewidth
  \else
    \Gin@nat@width
  \fi
}
\makeatother

\usepackage{thumbpdf,lmodern}
\usepackage{amsfonts,amsmath,amssymb,amsthm}
\usepackage{bbm,booktabs,setspace,longtable}
\usepackage[T1]{fontenc}

\newcommand{\fct}[1]{\code{#1()}}

\newcommand{\cA}{\mathcal{A}}
\newcommand{\cF}{\mathcal{F}}

\newcommand{\R}{\mathbb{R}}
\newcommand{\by}{\mathbf{y}}
\newcommand{\bx}{\mathbf{x}}
\newcommand{\bX}{\mathbf{X}}
\newcommand{\ind}{\mathbbm{1}}
\newcommand{\dd}{\hspace{0.1cm} \mathrm{d}}

\DeclareMathOperator{\CRPS}{CRPS}

\author{Sam Allen\\University of Bern\\Oeschger Centre for Climate Change Research}
\Plainauthor{Sam Allen}

\title{Weighted \pkg{scoringRules}: Emphasising Particular Outcomes when Evaluating Probabilistic Forecasts}
\Plaintitle{Weighted scoringRules: Emphasising Particular Outcomes when Evaluating Probabilistic Forecasts}
\Shorttitle{Weighted scoringRules}

\Abstract{
When predicting future events, it is common to issue forecasts that are probabilistic, in the form of probability distributions over the range of possible outcomes. Such forecasts can be evaluated using proper scoring rules. Proper scoring rules condense forecast performance into a single numerical value, allowing competing forecasters to be ranked and compared. To facilitate the use of scoring rules in practical applications, the \pkg{scoringRules} package in \proglang{R} provides popular scoring rules for a wide range of forecast distributions. This paper discusses an extension to the \pkg{scoringRules} package that additionally permits the implementation of popular weighted scoring rules. Weighted scoring rules allow particular outcomes to be targeted during forecast evaluation, recognising that certain outcomes are often of more interest than others when assessing forecast quality. This introduces the potential for very flexible, user-oriented evaluation of probabilistic forecasts. We discuss the theory underlying weighted scoring rules,
and describe how they can readily be implemented in practice using \pkg{scoringRules}. Functionality is available for weighted versions of several popular scoring rules, including the logarithmic score, the continuous ranked probability score (CRPS), and the energy score. Two case studies are presented to demonstrate this, whereby weighted scoring rules are applied to univariate and multivariate probabilistic forecasts in the fields of meteorology and economics.
}

\Plainkeywords{forecast evaluation, probabilistic forecasting, proper scoring rules, weighted scoring rules, R}

\Address{
  Sam Allen\\
  University of Bern\\
  Institute of Mathematical Statistics and Actuarial Science\\
  Alpeneggstrasse 22\\
  3012 Bern, Switzerland\\
  E-Mail: \email{sam.allen@unibe.ch}\\
  \emph{and}\\
  Oeschger Centre for Climate Change Research\\
}
\IfFileExists{upquote.sty}{\usepackage{upquote}}{}
\begin{document}

\section{Introduction: Weighted scoring rules}\label{sec:intro}

When predicting future events, it is common to issue forecasts that are probabilistic. Probabilistic forecasts generally take the form of probability distributions over the range of possible outcomes, comprehensively describing the predictive uncertainty. To assess the quality of a probabilistic forecast, scoring rules are functions $S(F, y)$ that take a forecast $F$ and the corresponding outcome $y$ as inputs, and output a numerical score that quantifies the forecast's accuracy. Scoring rules therefore condense forecast performance into a single value, providing a convenient framework with which to objectively rank and compare competing forecasts. As such, scoring rules have become a key component of probabilistic forecast evaluation.

To assess probabilistic forecasts in practice, the \pkg{scoringRules} package \citep{JordanEtAl2018} in the programming language \proglang{R} has become a widely used resource. The package contains analytical formulae for the two most popular univariate scoring rules --- the logarithmic score \citep[LogS;][]{Good1952} and the continuous ranked probability score \citep[CRPS;][]{MathesonWinkler1976} --- for forecast distributions belonging to a range of parametric families. These scoring rules are also available when the forecast is a sample from a predictive distribution, which is often the case in practice. The \pkg{scoringRules} package additionally allows samples from multivariate forecast distributions to be evaluated using popular multivariate scoring rules, including the energy score \citep{GneitingRaftery2007}, variogram score \citep{ScheuererHamill2015}, and a kernel score based on the Gaussian kernel \citep{GneitingRaftery2007}.

The scoring rules listed above assess forecasts made for all outcomes. While this is clearly desirable when assessing overall forecast performance, it is often the case that certain outcomes are of more interest than others. For example, one could argue that it is particularly important to issue accurate forecasts for outcomes that have a high impact on the forecast users. To emphasise particular outcomes during forecast evaluation, weighted scoring rules generalise conventional scoring rules by incorporating a weight function into the score. The weight function can be chosen such that a higher weight is assigned to outcomes that are of more interest. Weighted scoring rules therefore allow competing forecast systems to be ranked and compared when predicting particular outcomes, facilitating very flexible, user-oriented forecast evaluation.

Well known examples of weighted scoring rules include the conditional and censored likelihood scores proposed by \cite{DiksEtAl2011}, and the threshold-weighted CRPS introduced by \cite{MathesonWinkler1976} and \cite{GneitingRanjan2011}. However, the theory underlying weighted scoring rules extends beyond these examples: \cite{HolzmannKlar2017} demonstrate that the conditional and censored likelihood scores can be generalised to construct weighted versions of any proper scoring rule, while \cite{AllenEtAl2022} introduce a broad generalisation of the threshold-weighted CRPS that can be applied, for example, to probabilistic forecasts for multivariate outcomes.

In this paper, we describe how the \pkg{scoringRules} package has been extended to additionally permit the implementation of popular weighted scoring rules. While several alternative software packages exist to calculate particular scoring rules in certain situations \citep[see][for an overview]{JordanEtAl2018}, the development of weighted scoring rules is more recent. Until recently, for example, efficient application of popular weighted scoring rules was limited by theoretical considerations, leading to ad hoc implementations in practice \citep{SharpeEtAl2018}. Hence, to our knowledge, no other packages exist that provide a comprehensive collection of weighted scoring rules. We therefore hope that this extension to \pkg{scoringRules} will greatly facilitate the successful implementation of weighted scoring rules in practical applications.

In the following section, we review the existing theory of weighted scoring rules, and introduce examples of weighted versions of several popular scores, such as the LogS, the CRPS, and the energy score. The remainder of the paper then illustrates how these weighted scoring rules can be implemented in practice using the \pkg{scoringRules} package. Section \ref{sec:functionality} outlines the functionality of the package when calculating weighted scoring rules, and discusses implementation options. Section \ref{sec:examples} then presents two case studies in which these weighted scoring rules are used to target particular outcomes when evaluating probabilistic forecasts in practice. These case studies include applications in weather forecasting and economic forecasting, building on the examples presented in \cite{JordanEtAl2018}. A summary of the paper is presented in Section \ref{sec:summary}.

\section{Theoretical background}\label{sec:theory}

\subsection{Proper scoring rules}\label{sec:proper_scores}

Suppose we are interested in predicting a random variable $Y$ that takes values in a set $\Omega$, and that our forecasts are in the form of probability distributions over $\Omega$. Let $\cF$ denote a set of such forecasts. A scoring rule is a function
\begin{equation*}
	S: \cF \times \Omega \to \R \cup \{ -\infty, \infty \},
\end{equation*}
which takes a forecast $F \in \cF$ and an observation $y \in \Omega$ as inputs, and outputs a numerical value, or score, that quantifies the forecast accuracy. A lower score is assigned to a more accurate forecast. A scoring rule is proper with respect to $\cF$ if, when the observations are drawn from a distribution $G \in \cF$, the scoring rule is minimised in expectation by issuing $G$ as the forecast, i.e.
\begin{equation*}
	\E_{Y \sim G} S(G, Y) \leq \E_{Y \sim G} S(F, Y)
\end{equation*}
for all $F, G \in \cF$. If the above inequality is strict, then $S$ is strictly proper with respect to $\cF$.

When the outcome variable is real-valued ($\Omega \subseteq \R$), probabilistic forecasts are typically evaluated using either the logarithmic score (LogS) or the continuous ranked probability score (CRPS). The LogS is defined as
\begin{equation}\label{eq:logs}
	\mathrm{LogS}(F, y) = - \mathrm{log} f(y),
\end{equation}
where $f$ is the predictive density associated with the cumulative distribution function $F$ \citep{Good1952}. The CRPS is defined as
\begin{equation}\label{eq:crps}
\begin{split}
	\CRPS(F, y) &= \int_{\R} (F(z) - \ind\{ y \leq z\})^{2} \dd z \\
	&= \E_{F}|X - y| - \frac{1}{2} \E_{F} |X - X^{\prime}|,
\end{split}
\end{equation}
where $\ind$ is the indicator function, $X, X^{\prime} \sim F$ are independent random variables, and it is assumed in the second expression that $F$ has a finite mean \citep{MathesonWinkler1976,GneitingRaftery2007}. 

Generalisations of the LogS and the CRPS are also commonly used to evaluate probabilistic forecasts for multivariate outcomes, i.e. $\Omega \subseteq \R^{d}$ for $d > 1$. While the LogS in Equation \ref{eq:logs} can readily be applied to multivariate predictive densities, it is often the case that only a sample from the multivariate forecast distribution is available, making it difficult to employ the LogS in practice. Instead, alternative scoring rules have been proposed to evaluate multivariate probabilistic forecasts that can readily be applied to samples from a forecast distribution. 

Arguably the most well known multivariate scoring rule is the energy score \citep[ES;][]{GneitingRaftery2007}, which generalises the CRPS to higher dimensions:
\begin{equation}\label{eq:es}
	\mathrm{ES}(F, \by) = \E_{F} \| \bX - \by \| - \frac{1}{2} \E_{F} \| \bX - \bX^{\prime} \|,
\end{equation}
where $\|\cdot\|$ is the Euclidean distance in $\R^{d}$, $\by = (y_{1}, \dots, y_{d}) \in \Omega$, and $\bX = \left( X_{1}, \dots, X_{d} \right) $, $\bX^{\prime} = \left( X_{1}^{\prime}, \dots, X_{d}^{\prime} \right) \sim F$ are independent, with $F$ a probability distribution on $\Omega$. It is assumed here and throughout that the expectations are finite where necessary.

An alternative to the energy score is the variogram score \citep[VS;][]{ScheuererHamill2015}. The variogram score aims to explicitly assess the dependence structure of the multivariate forecast distributions by measuring the distance between the variogram of the forecast and that of the observation. The variogram score of order $p > 0$ is defined as
\begin{equation}\label{eq:vs}
	\mathrm{VS}^{p}(F, \by) = \sum_{i=1}^{d} \sum_{j=1}^{d} h_{i,j} \left( \E_{F} \left| X_{i} - X_{j} \right| ^{p} - \left| y_{i} - y_{j} \right| ^{p} \right)^{2},
\end{equation}
where $\bX = \left( X_{1}, \dots, X_{d} \right) \sim F$, $\by = \left( y_{1}, \dots, y_{d} \right) \in \Omega$, and $h_{i,j}$ are non-negative scaling parameters that control how much emphasis is given to a pair of dimensions. Following recommendations from \cite{ScheuererHamill2015}, the order of the score, $p$, is often chosen to be 0.5.

Both the energy score and the variogram score belong to the very general class of kernel scores \citep{GneitingRaftery2007}. Kernel scores are scoring rules that are constructed using conditionally negative definite kernels, and the kernel score framework has also been leveraged to introduce alternative multivariate scoring rules. \cite{AllenEtAl2022}, for example, introduced a multivariate scoring rule based on the inverse multiquadric kernel, while the so-called maximum mean discrepancy score (MMDS) is the kernel score corresponding to the Gaussian kernel:
\begin{equation}\label{eq:mmds}
	\mathrm{MMDS}(F, \by) = \frac{1}{2} \E_{F} \left[\mathrm{exp}\left\{ - \frac{1}{2} \| \bX - \bX^{\prime} \|^{2} \right\} \right] - \E_{F} \left[ \mathrm{exp}\left\{ - \frac{1}{2} \| \bX - \by \| ^{2} \right\} \right],
\end{equation}
where $\bX, \bX^{\prime} \sim F$ are independent.

To facilitate the implementation of these popular scoring rules in practice, the \pkg{scoringRules} package provides analytical expressions of the LogS and CRPS for forecasts that correspond to several familiar parametric distributions. It is also often the case that only a sample from the forecast distribution is available; this is common, for example, when considering ensemble forecasts issued by numerical weather and climate models, or output from Markov chain Monte Carlo (MCMC) algorithms \citep{KruegerEtAl2021}. The \pkg{scoringRules} package therefore additionally contains versions of the LogS, CRPS, ES, VS, and MMDS that can be used to evaluate forecasts in the form of a predictive sample. This can be achieved by replacing the expectations in Equations \ref{eq:crps}-\ref{eq:mmds} with sample means (see Appendix \ref{app:sampleformulas} for details). For the LogS, kernel density estimation is used to estimate the predictive density from the sample, prior to calculating the score.

\subsection{Weighted scoring rules}

The scoring rules introduced in the previous section evaluate the entire forecast distribution. However, one could argue that it is particularly important to issue accurate forecasts for events that have a high impact on the forecast users, and such events should therefore be given more weight during forecast evaluation. Weighted scoring rules achieve this by incorporating a non-negative weight function $w$ into conventional scoring rules, where the weight function determines how much emphasis should be placed on each possible outcome. Different approaches to weight scoring rules exist, and here we focus only on the two most popular frameworks.

\subsubsection{Outcome-weighted scoring rules}

\cite{DiksEtAl2011} introduced two weighted versions of the LogS that allow particular outcomes to be emphasised when calculating forecast accuracy. The conditional likelihood score (CoLS) is defined as 
\begin{equation*}
	\mathrm{CoLS}(F, y) = -w(y)\mathrm{log} f(y) + w(y) \mathrm{log} \left( \int_{\R} w(z) f(z) \dd z \right),
\end{equation*}
while the censored likelihood score (CeLS) is
\begin{equation*}
	\mathrm{CeLS}(F, y) = -w(y)\mathrm{log} f(y) - (1 - w(y)) \mathrm{log} \left(1 - \int_{\R} w(z) f(z) \dd z \right).
\end{equation*}
To understand how these weighted logarithmic scores behave, consider a weight function of the form $w(z) = \ind\{z \in \cA\}$, meaning only forecasts for outcomes in the set $ \cA \subseteq \R$ are of interest. In this example, if the observation $y \notin \cA$, then the CoLS is equal to zero, so that only the forecasts issued when $y \in \cA$ contribute to the score. If $y \in \cA$, then the CoLS is equivalent to the LogS applied to the conditional forecast distribution given that the observation is in $\cA$; forecast distributions are therefore assessed only via their restriction to the set $\cA$. The CeLS then extends the CoLS by additionally rewarding forecast distributions that can correctly predict when an outcome of interest will or will not occur. Note that if $\cA = \R$, then the weight function is always one, and both weighted scoring rules revert to the unweighted LogS.

\cite{HolzmannKlar2017} later generalised the CoLS and CeLS by demonstrating that this framework can readily be applied to any proper scoring rule. The resulting scoring rules, which we call outcome-weighted scoring rules, target particular outcomes by introducing a weighted version of the forecast distribution, and evaluating $F$ via its weighted representation. For the weight function $w(z) = \ind\{z \in \cA\}$, this weighted representation is simply the conditional distribution given that the outcome is in $\cA$, as discussed above for the CoLS and CeLS. Further details can be found in \cite{HolzmannKlar2017}.

An outcome-weighted CRPS can be defined as
\begin{equation*}
    \mathrm{owCRPS}(F, y) = \frac{1}{\bar{w}_{F}} \E_{F} \left[ |X - y|w(X)w(y) \right] - \frac{1}{2 \bar{w}_{F} ^{2}} \E_{F} \left[ |X - X^{\prime}|w(X)w(X^{\prime})w(y) \right],
\end{equation*}
where $X, X^{\prime} \sim F$ are independent and $\bar{w}_{F} = \E_{F} [w(X)]$.
Since this framework applies to any proper scoring rule, outcome-weighted versions of the ES, VS, and MMDS can similarly be introduced to target multivariate outcomes of interest during forecast evaluation:
\begin{equation*}
\begin{split}
	\mathrm{owES}(F, \by) =& \frac{1}{\bar{w}_{F}} \E_{F} \left[ \| \bX - \by \| w(\bX)w(\by) \right] - \frac{1}{2 \bar{w}_{F} ^{2}} \E_{F} \left[ \| \bX - \bX^{\prime} \| w(\bX)w(\bX^{\prime})w(\by) \right]; \\
	\mathrm{owVS}^{p}(F, \by) =& w(\by) \sum_{i=1}^{d} \sum_{j=1}^{d} h_{i,j} \left( \frac{1}{\bar{w}_{F}} \E_{F} \left[ \left| X_{i} - X_{j} \right| ^{p} w(\bX) \right] - \left| y_{i} - y_{j} \right| ^{p} \right)^{2}; \\
    \mathrm{owMMDS}(F, \by) =& \frac{1}{2 \bar{w}_{F}^{2}} \E_{F} \left[ \mathrm{exp}\left\{ - \frac{1}{2} \| \bX - \bX^{\prime} \| ^{2} \right\} w(\bX)w(\bX^{\prime})w(\by) \right] \\
    &- \frac{1}{\bar{w}_{F}} \E_{F} \left[ \mathrm{exp}\left\{ - \frac{1}{2} \| \bX - \by \| ^{2} \right\} w(\bX)w(\by) \right].
\end{split}
\end{equation*}

Note that in the multivariate case, the weight function takes a vector as an argument; $\bar{w}_{F}$ is thus defined as $\bar{w}_{F} = \E_{F} [w(\bX)]$.

The premise behind this class of weighted scoring rules is that, if attention is only on a particular set of outcomes, then the forecasts are only evaluated when these outcomes occur. When these outcomes do occur, the forecast distributions are evaluated using the conditional distribution given that the outcome of interest has occurred. In considering the conditional distribution given that an outcome of interest has occurred, the score does not consider the predicted probability that this outcome will occur. The CeLS extends the CoLS to address this, and suitable adaptations of the larger class of outcome-weighted scoring rules also exist, though these are not considered here \citep[see][]{HolzmannKlar2017}. 

Moreover, these scores are clearly not well-defined if the conditional distribution does not exist. This is equivalent to $\bar{w}_{F}$ being equal to zero, which could occur, for example, if $w(z) = \ind\{ z \in \cA \}$ and the forecast distribution assigns zero probability to the region $\cA$. The use of these outcome-weighted scoring rules is therefore only recommended when the weight function is strictly positive, or when interest is on events that are not rare, such that $\bar{w}_{F}$ is non-zero \citep{AllenEtAl2023}.

\subsubsection{Threshold-weighted scoring rules}

Arguably the most well known weighted scoring rule is the threshold-weighted CRPS proposed by \cite{MathesonWinkler1976} and \cite{GneitingRanjan2011}. The threshold-weighted CRPS introduces a weight function into the integral defining the CRPS:
\begin{equation*}
\begin{split}
	\mathrm{twCRPS}(F, y) &= \int_{\R} (F(z) - \ind\{ y \leq z \})^{2} w(z) \dd z \\
	&= \E_{F}|v(X) - v(y)| - \frac{1}{2} \E_{F}|v(X) - v(X^{\prime})|, 
\end{split}
\end{equation*}
where $v$ is any function such that $v(z) - v(z^\prime) = \int_{z^{\prime}}^{z} w(z) \dd z$ for all $z, z^{\prime} \in \R$ \citep{TaillardatEtAl2022,AllenEtAl2022}. We follow \cite{AllenEtAl2022} and refer to $v$ as a chaining function.

Just as we can generate outcome-weighted versions of any proper scoring rule, \cite{AllenEtAl2022} demonstrate that the theory underlying the threshold-weighted CRPS can readily be extended to any kernel score. As discussed, the ES, VS, and MMDS are all kernel scores, allowing threshold-weighted versions of these scores to be introduced:
\begin{equation}
\begin{split}
    \mathrm{twES}(F, \by) =& \E_{F} \| v(\bX) - v(\by) \| - \frac{1}{2} \E_{F} \| v(\bX) - v(\bX^{\prime}) \|; \\
    \mathrm{twVS}^{p}(F, \by) =& \sum_{i=1}^{d} \sum_{j=1}^{d} h_{i,j} \left( \E_{F} \left| v(\bX)_{i} - v(\bX)_{j} \right| ^{p} - \left| v(\by)_{i} - v(\by)_{j} \right| ^{p} \right)^{2}; \\
    \mathrm{twMMDS}(F, \by) =& \frac{1}{2} \E_{F} \left[\mathrm{exp}\left\{ - \frac{1}{2} \| v(\bX) - v(\bX^{\prime}) \| ^{2} \right\} \right] - \E_{F} \left[ \mathrm{exp}\left\{ - \frac{1}{2} \| v(\bX) - v(\by) \| ^{2} \right\} \right],
\end{split}
\end{equation}
where $\bX, \bX^{\prime} \sim F$ are independent random variables taking values on $\Omega \subseteq \R^{d}$, and $v: \R^{d} \to \R^{d}$ is a chaining function, so that $v(\by) = \left( v(\by)_{1}, \dots, v(\by)_{d} \right)$ and likewise for $v(\bX)$ and $v(\bX^{\prime})$.

In contrast to the outcome-weighted scoring rules, threshold-weighted scoring rules transform the forecasts and observations according to a chaining function $v$ prior to employing the unweighted version of the scores. The chaining function can therefore be chosen to focus the scoring rules on particular outcomes. While there exists a canonical way to obtain a chaining function from a given weight function in the univariate case, no such relationship exists when evaluating multivariate forecasts. This is discussed further in the following section.

\subsection{Weight and chaining functions}\label{sec:weight_chaining}

These weighted scoring rules provide attractive ways to target particular outcomes of interest when evaluating forecast performance, both in the univariate and multivariate case. In this section, we discuss possible weight and chaining functions that can be used within these weighted scoring rules. Certain choices can result in weighted scoring rules that are not proper, and these weight and chaining functions must therefore be chosen with care, to ensure that forecasters are not evaluated using an improper scoring rule. 

If both a weighted and unweighted version of a scoring rule are proper, then they will both be minimised on average by the same forecast distribution: the true distribution of the outcome. However, for two imperfect forecasts, the ranking of these forecasts may change depending on whether a weighted or unweighted scoring rule is employed. Weighted scoring rules may be less powerful than conventional scoring rules when discriminating between two forecast distributions, but they should be more discriminative when comparing forecasts made for particular outcomes. Put differently, if weighted scoring rules detect a difference between two forecast systems, then it is generally easier to interpret this difference than if it were detected using an unweighted scoring rule.

Readers are referred to \cite{GneitingRanjan2011}, \cite{LerchEtAl2017}, and \cite{AllenEtAl2022} for further details regarding what weight and chaining functions preserve the (strict) propriety of scoring rules. The weight and chaining functions that we consider here all result in weighted scoring rules that are themselves proper (though not necessarily strictly proper).

\subsubsection{Weight functions}

The choice of weight and chaining function is case-specific, and should depend on what information is to be extracted from the forecasts. Most commonly, interest is on outcomes within a certain range, or above or below a predefined threshold; this range or threshold may correspond to relevant quantiles of the previously observed outcomes. A univariate weight function that restricts attention to these events is
\begin{equation}\label{eq:uv_thr_weight}
    w(z) = \ind\{a < z < b\} \quad \text{for some} \quad - \infty \leq a < b \leq \infty,
\end{equation}
which is one if $z$ is between $a$ and $b$, and zero otherwise. To emphasise values above (below) some threshold $t$, we can set $a = t$ and $b = \infty$ ($a = -\infty$ and $b = t$).

Alternatively, certain events could be emphasised using a smoother weight function, which assigns a positive weight to all outcomes, but a higher weight to the events of interest. Popular weight functions to emphasise rare events include a Gaussian or logistic distribution function, e.g.
\begin{equation}\label{eq:gauss_weight}
    w(z) = \Phi_{\mu, \sigma} (z),
\end{equation}
where $\Phi_{\mu, \sigma}$ is the Gaussian distribution function with mean $\mu$ and standard deviation $\sigma$, with these parameters controlling the location of the weight function and the rate at which it tends to zero and one \citep{GneitingRanjan2011}. 

Gaussian and logistic density functions could additionally be used to target outcomes that are not rare. For example, the weight function 
\begin{equation*}
    w(z) = \phi_{\mu, \sigma}(z),
\end{equation*}
where $\phi_{\mu, \sigma}$ is the Gaussian density function with mean $\mu$ and standard deviation $\sigma$. This weight function will emphasise values around the location parameter $\mu$, with $\sigma$ determining the concentration of the weight around $\mu$.

Similar weight functions can also be used in the multivariate case. For example, it is common to define rare multivariate events as threshold exceedances that occur simultaneously in multiple dimensions, in which case a canonical weight function is
\begin{equation}\label{eq:mv_thr_weight}
    w(\mathbf{z}) = \ind\{a_{1} < z_{1} < b_{1}, \dots, a_{d} < z_{d} < b_{d} \} \quad \text{for} \quad - \infty \leq a_{i} < b_{i} \leq \infty, \quad i = 1, \dots, d.
\end{equation}
As in the univariate case, some values of the vectors $\mathbf{a} = (a_{1}, \dots, a_{d}) $ and $\mathbf{b} = (b_{1}, \dots, b_{d})$ can be set to $\pm \infty$ in order to focus on threshold exceedances. Multivariate Gaussian distribution and density functions could then again be used to target particular regions of multivariate space in a smoother way \citep{AllenEtAl2023}. These weight functions are listed in Table \ref{tab:weight_funcs}.

\renewcommand*{\arraystretch}{1.5}
\begin{table}
    \centering
    \begin{tabular}{l|l}
         Weight function & Chaining function \\
         \hline
         $w(z) = 1$ & $v(z) = z$ \\
         $w(z) = \ind\{ a < z < b \}$ & $v(z) = \min(\max(z, a), b) $ \\
         $w(z) = \Phi_{\mu, \sigma} (z)$ & $v(z) = (z - \mu) \Phi_{\mu, \sigma} (z) + \sigma^{2} \phi_{\mu, \sigma} (z)$ \\
         $w(z) = \phi_{\mu, \sigma} (z)$ & $v(z) = \Phi_{\mu, \sigma} (z)$ \\
         \hline 
         $w(\boldsymbol{z}) = \ind \left\{ a_{1} < z_{1} < b_{1}, \dots, a_{d} < z_{d} < b_{d} \right\}$ & $v(\boldsymbol{z})_{i} = \min(\max(z_{i}, a_{i}), b_{i})$ \\
         $w(\boldsymbol{z}) = \boldsymbol{\Phi_{\mu, \Sigma}} (\boldsymbol{z})$ & $v(\boldsymbol{z})_{i} = (z_{i} - \mu_{i}) \Phi_{\mu_{i}, \sigma_{i}} (z_{i}) + \sigma_{i} ^{2} \phi_{\mu_{i}, \sigma_{i}} (z_{i})$ \\
         $w(\boldsymbol{z}) = \boldsymbol{\phi_{\mu, \Sigma}} (\boldsymbol{z})$ & $v(\boldsymbol{z})_{i} = \Phi_{\mu_{i}, \sigma_{i}}(z_{i})$ \\
    \end{tabular}
    \caption{Examples of weight functions and chaining functions that could be used in weighted scoring rules. $\boldsymbol{\Phi_{\mu, \Sigma}}$ and $\boldsymbol{\phi_{\mu, \Sigma}}$ denote the multivariate Gaussian distribution and density functions with mean vector $\boldsymbol{\mu}=(\mu_{1}, \dots, \mu_{d})$ and covariance matrix $\Sigma$ with diagonal entries $\sigma_{1}, \dots, \sigma_{d}$. The multivariate chaining functions are component-wise extensions of the univariate chaining functions; hence, for concision, only the $i$-th component (for $i \in \{1, \dots, d\}$) of the multivariate chaining function is shown.} 
    \label{tab:weight_funcs}
\end{table}

\subsubsection{Chaining functions}

While the outcome-weighted scoring rules depend on a weight function, the threshold-weighted scoring rules depend on a chaining function. It is arguably less intuitive to choose a chaining function to emphasise certain outcomes of interest than a weight function. In the univariate case, the chaining function can be derived easily from a given weight function: we can take any function $v$ that satisfies
\begin{equation}\label{eq:univ_v}
    v(z) - v(z^\prime) = \int_{z^{\prime}}^{z} w(z) \dd z \quad \text{for all} \quad z, z^{\prime} \in \R.
\end{equation}
That is, $v$ is an anti-derivative of the chosen weight function. Table \ref{tab:weight_funcs} lists examples of chaining functions that correspond to the univariate weight functions given above.

\begin{figure}
\begin{Schunk}

{\centering \includegraphics[width=\maxwidth]{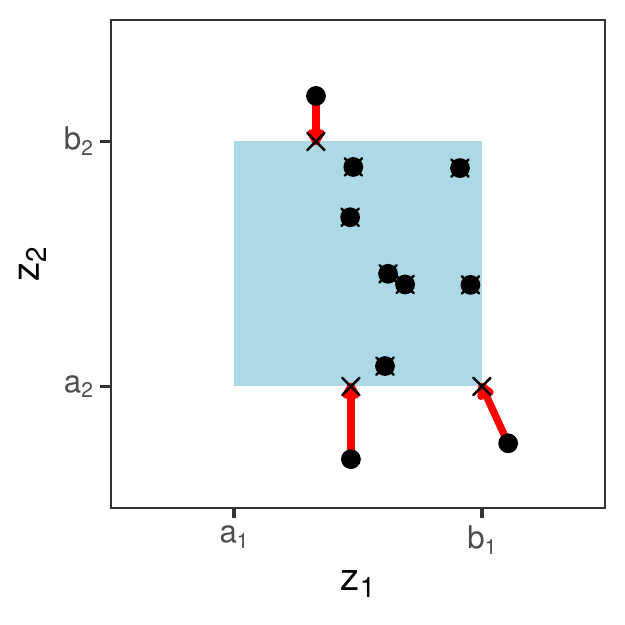} 

}

\end{Schunk}
\caption{Example of the weight function in Equation \ref{eq:mv_thr_weight} and the chaining function in Equation \ref{eq:default_v}. Ten two-dimensional observations are shown before (points) and after (crosses) applying the chaining function. The shaded region is the area specified by the weight function. Points in this region are unchanged after applying the chaining function, whereas points outside of this region are mapped onto the region's border, as indicated by the red arrows.}
\label{fig:chain_func}
\end{figure}

In the multivariate case, however, there is no canonical approach to derive a chaining function from a given weight function. \cite{AllenEtAl2022} discuss possible chaining functions that could be used to target certain multivariate outcomes when interest is on high-impact events. For the multivariate weight function in Equation \ref{eq:mv_thr_weight}, one possible chaining function is 
\begin{equation}\label{eq:default_v}
    v(z) = \left( \min(\max(z_{1}, a_{1}), b_{1}), \dots, \min(\max(z_{d}, a_{d}), b_{d}) \right),
\end{equation}
which is essentially a component-wise extension of the chaining function for the univariate weight in Equation \ref{eq:uv_thr_weight} (see Table \ref{tab:weight_funcs}). In this case, the weight function represents an orthant, or a box, in $\R^{d}$, and the chaining function projects points not in the orthant onto its perimeter; the points inside the orthant, i.e. for which the weight function is equal to one, remain unchanged. A two-dimensional example of this is given in Figure \ref{fig:chain_func}.

Similarly, for the smooth weight functions based on multivariate Gaussian distribution and density functions, a chaining function can be derived from a component-wise extension of the chaining functions corresponding to univariate Gaussian weight functions. Examples of such chaining functions are presented in Table \ref{tab:weight_funcs}. Note, however, that these component-wise extensions implicitly assume that the covariance matrix in the multivariate Gaussian weight function is diagonal. Readers are referred to \cite{AllenEtAl2022} for a more detailed discussion on multivariate chaining functions.

Although the weight and chaining functions presented in this section are simple examples that are frequently used in practice, the weighted scoring rules discussed herein can be employed with arbitrary such functions, permitting very flexible, user-oriented forecast evaluation. In the remainder of this paper, we will discuss how these weighted scoring rules have been integrated into the \pkg{scoringRules} package, facilitating their use in practical applications.

\section{Package functionality}\label{sec:functionality}

\subsection{Univariate weighted scoring rules}\label{sec:func_weighted_univ}

The weighted scoring rules discussed in the previous section can all be implemented using the \pkg{scoringRules} package. Functionality is currently available for probabilistic forecasts that take the form of a predictive sample. In this case, it is straightforward to calculate the weighted scoring rules with arbitrary, user-specified weight functions, which is generally not the case for parametric families of distributions. Expressions for the weighted scoring rules discussed in the previous section when the forecast is a predictive sample are given in Appendix \ref{app:sampleformulas}.

The \pkg{scoringRules} package already contains functions to calculate the LogS, CRPS, ES, VS, and MMDS for forecasts in the form of predictive samples. Suppose the sample is comprised of $m$ members. As explained in \cite{JordanEtAl2018}, the naming convention of these functions is \fct{[score]\_sample}, where \code{[score]} refers to the scoring rule to be calculated. These functions take the observed value(s) and the forecast samples as inputs, and output the desired score value. For example, to calculate the CRPS corresponding to a vector of $n$ observations \code{y} and a $n \times m$ matrix \code{dat} whose rows contain the $m$ forecast samples corresponding to each observation, one could use
\begin{Code}
crps_sample(y, dat)
\end{Code}
The output is a numeric vector containing the score for each of the $n$ forecast cases.

The same convention is adopted for the weighted scoring rules. In the univariate case, the following functions calculate the outcome-weighted and threshold-weighted CRPS, and the conditional or censored likelihood scores:
\begin{Code}
owcrps_sample(y, dat, a = -Inf, b = Inf, 
              weight_func = function(x) as.numeric(x > a & x < b),
              w = NULL, show_messages = TRUE)
twcrps_sample(y, dat, a = -Inf, b = Inf, 
              chain_func = function(x) pmin(pmax(x, a), b),
              w = NULL, show_messages = TRUE)
clogs_sample(y, dat, a = -Inf, b = Inf, bw = NULL, 
             show_messages = FALSE, cens = TRUE)
\end{Code}
The \code{cens} argument in \code{clogs_sample} specifies whether the conditional likelihood score or the censored likelihood score should be returned; the default is \code{cens = TRUE}, in which case the CeLS is calculated. 

As discussed in Section \ref{sec:proper_scores}, the LogS takes a predictive density as input, and hence cannot readily be applied to predictive samples. To circumvent this, \code{logs_sample} employs kernel density estimation to estimate a predictive density from the sample, and then calculates the LogS from the estimated density function. However, \cite{KruegerEtAl2021} demonstrate that the resulting score is sensitive to the bandwidth parameter \code{bw} of the kernel density estimation, and the authors therefore recommended using the CRPS instead of the LogS, particularly when the sample size $m$ is small. Similarly, the conditional and censored likelihood scores also require a predictive density as inputs, and kernel density estimation is used to estimate this from the predictive sample prior to calculating the weighted scores. We anticipate that these weighted scores will be yet more sensitive to the kernel density estimation parameters, especially when a weight function is used that targets more extreme outcomes. As such, when the forecast is in the form of a predictive sample, we similarly recommend employing weighted versions of the CRPS, rather than the conditional or censored likelihood score.

In addition to observations and forecast samples, the functions listed above have arguments that allow particular outcomes to be targeted when calculating the weighted scores. By default, the weighted scoring rules employ the weight function $w(z) = \ind\{a < z < b\}$, which, as discussed in the previous section, is most commonly applied in practice. The arguments \code{a} and \code{b} are single numeric values representing the lower and upper bounds in this weight function, respectively. If these arguments are not specified, then their default values are \code{a = -Inf} and \code{b = Inf}, resulting in a weight function that is always one, and thus recovering the unweighted scoring rules.

\begin{Schunk}
\begin{Sinput}
R> obs <- rnorm(5)
R> sample_m <- matrix(rnorm(5e4), nrow = 5)
R> score_df <- data.frame(crps = crps_sample(obs, sample_m),
+                         owcrps = owcrps_sample(obs, sample_m),
+                         twcrps = twcrps_sample(obs, sample_m))
R> print(score_df)
\end{Sinput}
\begin{Soutput}
   crps owcrps twcrps
1 0.275  0.275  0.275
2 1.230  1.230  1.230
3 0.246  0.246  0.246
4 0.764  0.764  0.764
5 1.355  1.355  1.355
\end{Soutput}
\end{Schunk}

On the other hand, if we want to emphasise outcomes above a threshold \code{t}, then we can set the lower bound in the weight function to \code{a = t}, and the upper bound to \code{b = Inf}.

\begin{Schunk}
\begin{Sinput}
R> t <- 0
R> score_df <- data.frame(crps = crps_sample(obs, sample_m),
+                         owcrps = owcrps_sample(obs, sample_m, a = t),
+                         twcrps = twcrps_sample(obs, sample_m, a = t))
R> print(score_df)
\end{Sinput}
\begin{Soutput}
   crps owcrps twcrps
1 0.275  0.000  0.120
2 1.230  0.000  0.115
3 0.246  0.000  0.119
4 0.764  0.306  0.645
5 1.355  0.809  1.235
\end{Soutput}
\end{Schunk}

Similarly, if we want to emphasise values below the threshold, then we can set \code{a = -Inf} and \code{b = t}. To avoid misuse, an error is returned if \code{a} is not smaller than \code{b}.

\begin{figure}	
\begin{Schunk}

{\centering \includegraphics[width=\linewidth]{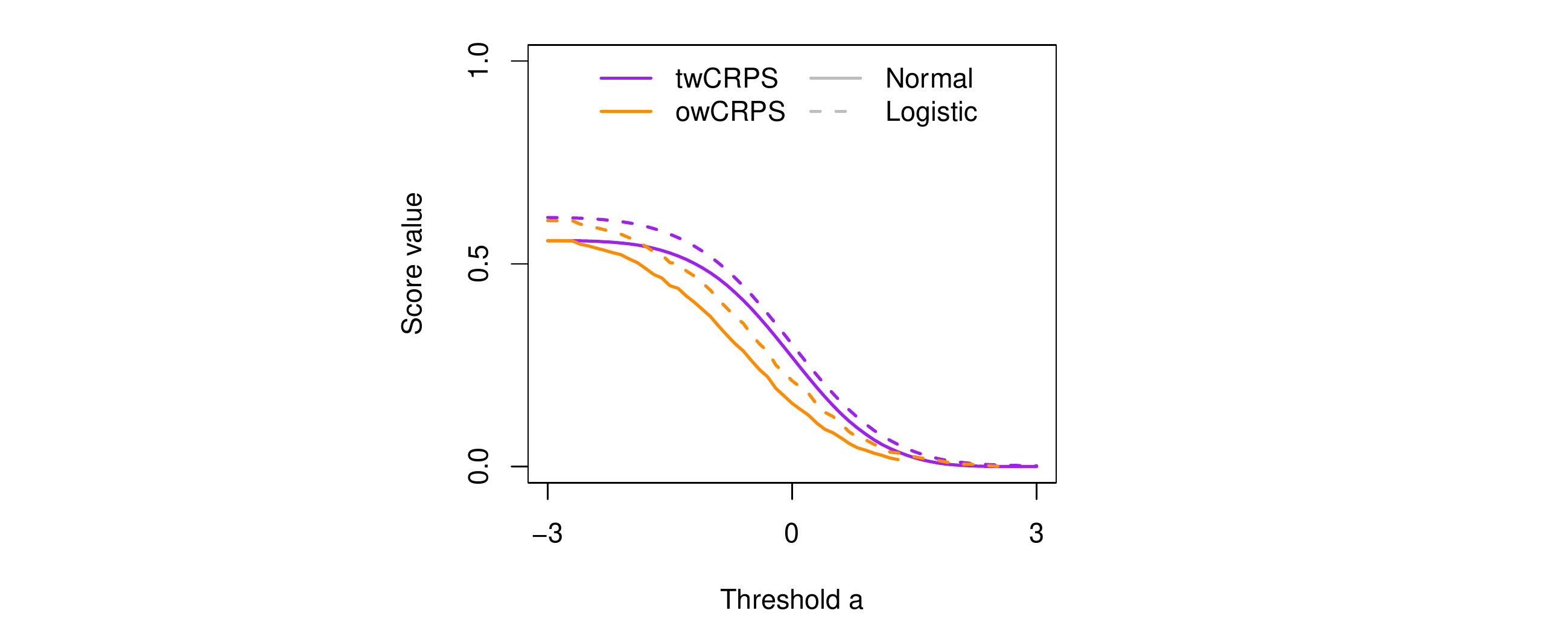} 

}

\end{Schunk}
\caption{Average twCRPS (purple) and owCRPS (orange) as a function of the threshold $a$ in the weight function $w(z) = \ind\{z > a\}$. The observations are drawn from a standard normal distribution, and scores are shown for predictive samples from a standard normal (solid) and standard logistic (dashed) distribution. Note that for high thresholds, the owCRPS is not always well-defined.}
\label{fig:score_vs_t}
\end{figure}

A useful diagnostic tool is to plot the average score as a function of the threshold. In this case, as the lower bound in the weight function \code{a} becomes smaller (or the upper bound \code{b} becomes larger), the weighted score tends to the unweighted score, allowing the user to simultaneously visualise overall forecast performance, as well as performance when predicting particular outcomes \citep{GneitingRanjan2011}. An example of this is presented in Figure \ref{fig:score_vs_t}, where the outcome-weighted CRPS and threshold-weighted CRPS for two forecasts distributions are displayed as a function of \code{a} in the default weight function, with \code{b = Inf}.

\subsection{Multivariate weighted scoring rules}\label{sec:func_weighted_multiv}

Similarly to the LogS and CRPS, \pkg{scoringRules} contains functions to calculate the ES, VS, and MMDS for multivariate forecast distributions in the form of predictive samples.
\begin{Code}
es_sample(y, dat, w = NULL)
vs_sample(y, dat, w = NULL, w_vs = NULL, p = 0.5)
mmds_sample(y, dat, w = NULL)
\end{Code}

These multivariate scoring rule functions can only evaluate a single multivariate forecast at a time. Hence, the observation argument \code{y} is a vector of length $d$, representing an element in $\R^{d}$, the forecast argument \code{dat} is a $d \times m$ matrix, with the columns representing the simulated samples (or ensemble members) from the multivariate forecast distribution, and the output is a single value. These functions can then be sequentially applied to multiple forecast cases using the \fct{apply} functions or \code{for} loops \citep[see Appendix B of][]{JordanEtAl2018}.

Similarly, outcome-weighted and threshold-weighted versions of these multivariate scoring rules are calculated using
\begin{Code}
owes_sample(y, dat, a = -Inf, b = Inf,
            weight_func = function(x) as.numeric(all(x > a & x < b)), 
            w = NULL)
owvs_sample(y, dat, a = -Inf, b = Inf,
            weight_func = function(x) as.numeric(all(x > a & x < b)),
            w = NULL, w_vs = NULL, p = 0.5)
owmmds_sample(y, dat, a = -Inf, b = Inf,
              weight_func = function(x) as.numeric(all(x > a & x < b)), 
              w = NULL)      
twes_sample(y, dat, a = -Inf, b = Inf,
            chain_func = function(x) pmin(pmax(x, a), b), w = NULL)
twvs_sample(y, dat, a = -Inf, b = Inf,
            chain_func = function(x) pmin(pmax(x, a), b),
            w = NULL, w_vs = NULL, p = 0.5)
twmmds_sample(y, dat, a = -Inf, b = Inf,
              chain_func = function(x) pmin(pmax(x, a), b), w = NULL)
\end{Code}

As in the univariate case, the default weight function corresponds to Equation \ref{eq:mv_thr_weight}, where interest is on a range of values in each dimension. The default chaining function used to calculate the threshold-weighted scores is Equation \ref{eq:default_v}. Arguments \code{a} and \code{b} are again used to define the lower and upper bounds of the default weight function. In contrast to the univariate case, however, \code{a} and \code{b} are numeric vectors of length $d$, rather than single values. 

If the input value of \code{a} or \code{b} is a single value, then it is automatically converted into a vector of length $d$, all containing the same element. The default values are \code{a = -Inf} and \code{b = Inf}, which again returns the unweighted scoring rule. Hence, if we want to emphasise values above the same threshold $t$ in all dimensions, then we could either use \code{a = c(t, t, ...)} and \code{b = c(Inf, Inf, ...)}, or we could use \code{a = t} and \code{b = Inf}. For example, for the threshold-weighted energy score, we have 
\begin{Schunk}
\begin{Sinput}
R> d <- length(obs)
R> twes_sample(obs, sample_m, a = t)
\end{Sinput}
\begin{Soutput}
[1] 1.34
\end{Soutput}
\begin{Sinput}
R> twes_sample(obs, sample_m, a = rep(t, d))
\end{Sinput}
\begin{Soutput}
[1] 1.34
\end{Soutput}
\end{Schunk}

Finally, note that the functions to calculate the multivariate weighted scores also include optional weight arguments that cannot be used to target particular outcomes of interest. The argument \code{w} is a vector of length $m$ that allows more weight to be given to particular elements of the sample in the forecast distribution. This argument is also available when calculating the unweighted scoring rules, and the univariate weighted scores. The variogram score functions additionally have an argument \code{w_vs}, which is a $d \times d$ matrix containing the scaling parameters $h_{i,j}$ in Equation \ref{eq:vs}. These scaling parameters put more emphasis on combinations of dimensions of the multivariate variables, rather than targeting particular outcomes.

\subsection{Custom weight and chaining functions}\label{sec:custom_weight}

The functions to calculate the weighted scoring rules use a default weight function that assumes emphasis is to be placed on a particular region of the outcome space. Although this weight function is frequently applied in practice, it may be the case that another weight function is desired. As discussed, the motivation for considering only forecasts in the form of a predictive sample is that it is straightforward to calculate the resulting scores for arbitrary weight and chaining functions. The weighted scoring rule functions in \pkg{scoringRules} therefore additionally contain an argument that allows for a custom weight or chaining function to be used.

The \code{weight_func} argument can be used to incorporate a custom weight function into the outcome-weighted scoring rules. This argument must be a function that takes a vector as an input, and outputs either a vector of the same length as the input (if a univariate scoring rule is being used), or a single numeric value (if a multivariate scoring rule is used). An error is returned if the weight function is found to return negative weights, or if the output is not of the correct format.

For example, consider the Gaussian distribution function in Equation \ref{eq:gauss_weight}, with location parameter \code{mu} and scale parameter \code{sigma}. To use this as the weight function when calculating the outcome-weighted CRPS, one could use

\begin{Schunk}
\begin{Sinput}
R> mu <- 0; sigma <- 1
R> weight_func <- function(x) pnorm(x, mean = mu, sd = sigma)
R> owcrps_sample(obs, sample_m, weight_func = weight_func)
\end{Sinput}
\begin{Soutput}
[1] 0.2002 0.0703 0.1868 0.3524 0.8788
\end{Soutput}
\end{Schunk}

Similarly, a multivariate Gaussian distribution could be used as a multivariate weight function. Let \code{mu} be the mean vector of this distribution, and assume the covariance matrix is diagonal with entries $\sigma_{1}, \dots, \sigma_{d}$. Then, the outcome-weighted ES with this weight function can be calculated using

\begin{Schunk}
\begin{Sinput}
R> mu <- rnorm(d, 0, 0.5); sigma <- runif(d, 0.5, 1.5)
R> weight_func <- function(x) prod(pnorm(x, mean = mu, sd = sigma))
R> owes_sample(obs, sample_m, weight_func = weight_func)
\end{Sinput}
\begin{Soutput}
[1] 0.0418
\end{Soutput}
\end{Schunk}

Whereas the outcome-weighted scores depend on a weight function, the threshold-weighted scores rely on a chaining function. For the threshold-weighted CRPS, a chaining function corresponds directly to a weight function via Equation \ref{eq:univ_v}. However, computation of the threshold-weighted CRPS for a sample forecast requires the chaining function rather than a weight function, and hence functionality is not currently available to take a weight function as an argument. In this case, it is necessary to derive the chaining function corresponding to the weight. For the simple weight functions commonly used in practice, this is typically straightforward to achieve (see Table \ref{tab:weight_funcs} for popular choices).

The \code{chain_func} argument can be used to incorporate a custom chaining function into the threshold-weighted scoring rules. In contrast to \code{weight_func}, the \code{chain_func} argument should be a function whose inputs and outputs are the same length as the observation input \code{y}. For example, in the multivariate case, this function should both input and output a vector of length $d$.

In the univariate case, if the chaining function satisfies Equation \ref{eq:univ_v} for some non-negative weight function $w$, then it will be a non-decreasing function; that is, if $z > z^{\prime}$, then $v(z) \geq v(z^{\prime})$ for all $z, z^{\prime} \in \R$. While a decreasing chaining function could also be used within Equation \ref{eq:crps}, this does not correspond to the original definition of the twCRPS presented in \cite{GneitingRanjan2011}, and is therefore not recommended: a warning message is returned if \code{chain_func} is found to be decreasing.

Table \ref{tab:weight_funcs} contains possible chaining functions corresponding to the Gaussian weight functions employed above. These chaining functions can be implemented within \code{twcrps_sample} and \code{twes_sample} as follows

\begin{Schunk}
\begin{Sinput}
R> chain_func <- function(x) (x - mu)*pnorm(x, mu, sigma) + 
+    (sigma^2)*dnorm(x, mu, sigma)
R> mu <- 0; sigma <- 1
R> twcrps_sample(obs, sample_m, chain_func = chain_func)
\end{Sinput}
\begin{Soutput}
[1] 0.135 0.263 0.123 0.528 1.082
\end{Soutput}
\begin{Sinput}
R> mu <- rnorm(d, 0, 0.5); sigma <- runif(d, 0.5, 1.5)
R> twes_sample(obs, sample_m, chain_func = chain_func)
\end{Sinput}
\begin{Soutput}
[1] 1.48
\end{Soutput}
\end{Schunk}
Weighted versions of the other scoring rules discussed herein can be calculated analogously.

It is challenging to construct general analytical formulae for weighted scoring rules corresponding to parametric forecast distributions and arbitrary weight functions. Similarly, since the CoLS and CeLS require kernel density estimation to estimate the predictive density given the sample, these scores cannot be readily implemented with arbitrary weight functions. Hence, \code{clogs_sample} does not take custom weight or chaining functions as arguments, so that only the default weight function is available for these scores.

\section{Usage examples}\label{sec:examples}

\cite{JordanEtAl2018} present two practical applications in which the \pkg{scoringRules} functionality is used to evaluate probabilistic forecasts. In this section, we revisit these applications, and illustrate how the weighted scoring rules available in \pkg{scoringRules} allow particular outcomes to be targeted during forecast evaluation. In both examples, the data and probabilistic models are as described in \cite{JordanEtAl2018}, and further details can be found therein.

\subsection{Probabilistic weather forecasting via ensemble post-processing}
Firstly, consider forecasts of precipitation accumulation in Innsbruck, Austria. The \code{RainIbk} data set in the \pkg{crch} \proglang{R} package \citep{MessnerEtAl2016} contains three-day precipitation accumulations recorded in Innsbruck from January 2000 to September 2013. Forecasts for these precipitation accumulations can be obtained from numerical weather prediction (NWP) models, which use physical laws to emulate the evolution of the atmosphere through time. These models are typically run several times, using different initial conditions and possibly different model configurations, yielding an ensemble of predictions that characterises the forecast uncertainty. The \code{RainIbk} data set contains 11-member ensemble forecasts corresponding to the three-day precipitation accumulations between five and eight days in advance.

However, operational ensemble forecasts tend to exhibit systematic errors when predicting surface weather variables such as precipitation. Hence, it is common for the ensemble forecasts to undergo some form of statistical post-processing. Post-processing methods try to learn the systematic errors that manifest in the NWP models, and then remove them from the forecasts. While a number of statistical post-processing methods have been proposed, the methods considered here assume that the square root of the precipitation accumulation follows a parametric distribution. The location and scale parameters of this distribution are assumed to depend linearly on the mean and the log-transformed standard deviation of the ensemble members, respectively. This general framework for post-processing ensemble weather forecasts is typically known as non-homogeneous regression or ensemble model output statistics \citep[EMOS;][]{GneitingEtAl2005}. 

Three alternative parametric distributions are then compared within this framework: a logistic, Gaussian, and Student's $t$ distribution. These three parametric distributions are censored below at zero, resulting in a forecast distribution that assigns zero probability to negative precipitation accumulations, and a non-negligible probability to zero precipitation. Further details about the statistical post-processing methods are available in \cite{JordanEtAl2018} and references therein.

Data from January 2000 to November 2004 is used to train the statistical post-processing models, and the resulting forecasts are then evaluated out-of-sample using the data from January 2005 to September 2013. The models are fit to the training data using maximum likelihood estimation via the \pkg{crch} package. Applying these models to the ensemble forecasts in the test data set returns predictive location and scale parameters for each model and each forecast case. For concision, we only show the code used to evaluate the Gaussian forecast distributions; this can easily be extended to the other models. In this case, the vectors \code{gauss_mu} and \code{gauss_sc} contain the estimated location and scale parameters that characterise the predictive distributions in the test data, while \code{obs} represents the time series of the corresponding observed precipitation accumulations. 

These three post-processing models can then be evaluated and compared using scoring rules. \cite{JordanEtAl2018} demonstrate how the CRPS can be used for this purpose. 

\begin{Schunk}
\begin{Sinput}
R> gauss_crps <- crps_cnorm(y = obs, location = gauss_mu, scale = gauss_sc, 
+                           lower = 0, upper = Inf)
\end{Sinput}
\end{Schunk}

The result is a vector of scores corresponding to each forecast case in the test data set, and the competing forecast strategies can then be compared using their average scores.

\begin{Schunk}
\begin{Sinput}
R> scores <- data.frame(Logistic = logis_crps, Gaussian = gauss_crps,
+                       Students_t = stud_crps, Ensemble = ens_crps)
R> sapply(scores, mean)
\end{Sinput}
\begin{Soutput}
  Logistic   Gaussian Students_t   Ensemble 
     0.875      0.876      0.875      1.321 
\end{Soutput}
\end{Schunk}

The mean CRPS values indicate that all post-processing models substantially improve upon the raw ensemble forecasts, and there are only small differences between the post-processing models. Of course, in a formal study, we should accompany these scores with measures of uncertainty, or perform statistical tests that clarify whether the differences between the scores are significant.

While the CRPS assesses overall forecast performance, the three post-processing models could also be compared with respect to their predictions of particular outcomes. This can be achieved here using weighted versions of the CRPS. As discussed, weighted scoring rules are only available in \pkg{scoringRules} for forecasts in the form of a predictive sample. Hence, to evaluate the post-processed forecast distributions, we must first sample from the predictive distributions, and use this sample to estimate the score for the parametric forecasts. Here, we sample 1000 observations from the predictive distributions, which should provide a reasonable approximation of the score for the continuous forecast distribution \citep[Figure 2]{JordanEtAl2018}.

\begin{Schunk}
\begin{Sinput}
R> ens_size <- 1000
R> n <- length(obs)
R> gauss_sample <- replicate(ens_size, rnorm(n, gauss_mu, gauss_sc))
R> gauss_sample[gauss_sample < 0] <- 0
\end{Sinput}
\end{Schunk}

An obvious question concerns which weight function to employ within the weighted scores. Weight and chaining functions should be chosen such that the outcomes that are of most interest to the practitioners are emphasised. As discussed, this will likely change on a case-by-case basis, and more than one weight function could be employed to gain a more complete understanding of the forecast performance; summarising plots such as Figure \ref{fig:score_vs_t} are particularly useful.

In a weather forecasting context, the relationship between weather conditions and socio-economical impacts is relatively well understood. Hence, an obvious choice of the weight function is one that reflects the costs associated with each possible outcome. Outcomes that correspond to higher impacts would therefore be emphasised in the weighted scores, and forecasters would be encouraged to issue more accurate forecasts for these high-impact events. Alternatively, to reduce the impact of these events, national weather centres issue warnings to the general public. These warnings typically correspond to relevant thresholds of the outcome variable, determined from climatological records and thorough analyses of the risks of weather to infrastructure and public health. A simple alternative weight function would thus use an indicator weight function (e.g. Equation \ref{eq:uv_thr_weight}), with the parameters in this weight function defined by the warning thresholds.

For the rainfall forecasts considered here, we firstly employ this latter weight function to emphasise values above a threshold of interest $t$, i.e. $w(z) = \ind\{ z > t \}$. This weight function can be employed in \pkg{scoringRules} by setting the arguments \code{a = t} and \code{b = Inf} in the weighted scoring rule functions. In doing so, the weighted scoring rules will assess the forecasts in their ability to predict high precipitation accumulations, which are of particular relevance since they often lead to flooding. As a threshold, we choose $t=\sqrt{30}$mm. This choice was made since $30$mm is commonly chosen as a threshold in rainfall warning systems, and $\sqrt{30}$ is roughly equal to the 95th percentile of the square root-transformed precipitation values in the training data considered here.

\begin{Schunk}
\begin{Sinput}
R> t <- sqrt(30)
R> gauss_twcrps <- twcrps_sample(y = obs, dat = gauss_sample, a = t)
\end{Sinput}
\end{Schunk}

Having repeated this for the logistic and Student's $t$ forecast distributions, we can then compare the three post-processing models using their average threshold-weighted CRPS.

\begin{Schunk}
\begin{Sinput}
R> scores <- data.frame(Logistic = logis_twcrps, Gaussian = gauss_twcrps,
+                       Students_t = stud_twcrps, Ensemble = ens_twcrps)
R> sapply(scores, mean)
\end{Sinput}
\begin{Soutput}
  Logistic   Gaussian Students_t   Ensemble 
    0.0491     0.0490     0.0489     0.0774 
\end{Soutput}
\end{Schunk}

These threshold-weighted CRPS values illustrate that, while the three post-processing methods are again almost indistinguishable, the relative improvement of the post-processed forecasts upon the raw ensemble forecasts has increased. This suggests that post-processing is particularly beneficial when predicting more extreme precipitation accumulations.  

We could also calculate the outcome-weighted CRPS for these forecasts in a similar way. However, when interest is on extreme events, there is a chance that neither the observation nor any members of the predictive sample exceed the threshold of interest, resulting in an undefined outcome-weighted score.

Alternatively, the Gaussian distribution function could also be used to emphasise larger precipitation values without restricting attention only to values above a threshold. The threshold-weighted CRPS values corresponding to this weight function are given below.
The results largely agree with those observed for the previous weight function. 

\begin{Schunk}
\begin{Sinput}
R> sigma <- 1
R> weight_func <- function(x) pnorm(x, mean = t, sd = sigma)
R> chain_func <- function(x) (x - t)*pnorm(x, mean = t, sd = sigma) + 
+    (sigma^2)*dnorm(x, mean = t, sd = sigma)
R> gauss_twcrps <- twcrps_sample(obs, gauss_sample, chain_func = chain_func)
\end{Sinput}
\end{Schunk}

\begin{Schunk}
\begin{Sinput}
R> scores <- data.frame(Logistic = logis_twcrps, Gaussian = gauss_twcrps,
+                       Students_t = stud_twcrps, Ensemble = ens_twcrps)
R> sapply(scores, mean)
\end{Sinput}
\begin{Soutput}
  Logistic   Gaussian Students_t   Ensemble 
    0.0676     0.0676     0.0674     0.1079 
\end{Soutput}
\end{Schunk}

The choice of \code{sigma} above is somewhat arbitrary, and depends on how quickly the weight function should tend to zero or one. If \code{sigma} is equal to zero, then, in theory, we should recover the indicator weight function employed above.

In general, the performance of the forecasts will change depending on the weight function. The weight functions discussed herein are only examples, and the choice of weight function will depend on the application. Other weight functions could also readily be employed. 

\subsection{Bayesian forecasts of US GDP growth}

The second case study considers an example from economics. It is standard for national banks to issue forecasts for the country's gross domestic product (GDP) growth in the coming quarters. In this example, probabilistic forecasts of US GDP growth (in \%) are obtained using a Markov switching autoregressive model, with exact details given in \cite{KruegerEtAl2021}. This model is used to derive forecasts of quarterly US GDP growth for the following year, i.e. the next four quarters.

The data used in this example is available from the data set \code{gdp} in \pkg{scoringRules}, which contains observed US GDP growth for 271 quarters between 1947 and 2014. We use the data prior to 2014 as training data to estimate the Markov switching autoregressive model, which is then used to forecast the GDP growth in the four quarters of 2014.

The model implemented here is Bayesian, and is estimated using Markov chain Monte Carlo (MCMC) methods. As is common for Bayesian models that employ MCMC, the analytical form of the predictive distribution is not known. The forecast distributions considered here are therefore predictive samples obtained from the MCMC algorithm. The resulting forecast distributions are displayed in Figure 4 of \cite{JordanEtAl2018}. 

These forecast distributions for each quarter can be evaluated univariately using both the CRPS and the LogS. In the following, \code{obs} denotes a vector containing the four observed GDP growths in 2014, while \code{X} is a matrix containing the MCMC predictions. 

\begin{Schunk}
\begin{Sinput}
R> scores_crps <- crps_sample(obs, X)
R> scores_logs <- logs_sample(obs, X)
R> print(cbind(scores_crps, scores_logs))
\end{Sinput}
\begin{Soutput}
       scores_crps scores_logs
2014Q1       3.454        4.05
2014Q2       1.322        2.25
2014Q3       1.711        2.51
2014Q4       0.719        1.96
\end{Soutput}
\end{Schunk}

As in the previous example, weighted versions of these scoring rules can again be used to target particular outcomes when quantifying forecast performance. The choice of weight function will again depend on what information is most relevant for practitioners. When forecasting GDP growth, it is particularly important to accurately predict when growth rates will be negative, since this suggests a decline in the country's economy. To emphasise negative growth rates within weighted scoring rules, a canonical weight function is $w(z) = \ind\{z < 0 \}$. Outcome-weighted and threshold-weighted CRPS values, as well as conditional and censored likelihood scores, are shown below for the Markov switching autoregressive forecasts considered here.

\begin{Schunk}
\begin{Sinput}
R> t <- 0
R> scores_owcrps <- owcrps_sample(obs, X, b = t)
R> scores_twcrps <- twcrps_sample(obs, X, b = t)
R> scores_cols <- clogs_sample(obs, X, b = t, cens = FALSE)
R> scores_cels <- clogs_sample(obs, X, b = t)
R> print(cbind(scores_owcrps, scores_twcrps, scores_cols, scores_cels))
\end{Sinput}
\begin{Soutput}
       scores_owcrps scores_twcrps scores_cols scores_cels
2014Q1         0.653        1.8714        1.97       4.049
2014Q2         0.000        0.0303        0.00       0.210
2014Q3         0.000        0.0501        0.00       0.258
2014Q4         0.000        0.0586        0.00       0.278
\end{Soutput}
\end{Schunk}

The outcome-weighted CRPS and the conditional likelihood scores are zero for the last three quarters, since the observed GDP growths are greater than zero in these cases. The threshold-weighted CRPS and censored likelihood score, on the other hand, additionally assess the forecast probability that is assigned to a positive GDP growth occurring. 

One could argue that economists also receive considerable attention when exceptionally high growth is forecast. To address this, a weight function could be used that simultaneously emphasises low and high GDP growth: e.g. $w(z) = \ind\{ z < 0 \} + \ind\{ z > t \}$ for some reasonably high threshold $t > 0$. Although this does not align with the default weight function used within \pkg{scoringRules}, the \code{weight_func} and \code{chain_func} arguments allow this custom weight function to be employed within the weighted versions of the CRPS. The resulting scores are presented below. In this case, $t$ is chosen to be 9\%, which again roughly corresponds to the 95th percentile of the previously observed GDP growths.

\begin{Schunk}
\begin{Sinput}
R> a <- 0
R> b <- 9 
R> weight_func <- function(x) as.numeric((x < a) | (x > b))
R> chain_func <- function(x) (x < a)*(x - a) + (x > b)*(x - b) + a
\end{Sinput}
\end{Schunk}

\begin{Schunk}
\begin{Sinput}
R> scores_owcrps <- owcrps_sample(obs, X, weight_func = weight_func)
R> scores_twcrps <- twcrps_sample(obs, X, chain_func = chain_func)
R> print(cbind(scores_owcrps, scores_twcrps))
\end{Sinput}
\begin{Soutput}
       scores_owcrps scores_twcrps
2014Q1         0.739        1.8715
2014Q2         0.000        0.0306
2014Q3         0.000        0.0504
2014Q4         0.000        0.0592
\end{Soutput}
\end{Schunk}

Yet more relevant than forecasting negative GDP growth in an individual quarter is predicting a decline in GDP growth in successive quarters; this is commonly used by analysts as an indicator for a recession \citep{BEA2023}. Hence, utilising this definition, evaluating forecasts for recessions becomes a multivariate problem. 

The weighted multivariate scoring rules discussed herein allow us to emphasise successive quarters with negative GDP growth when evaluating forecast accuracy. If we consider two consecutive declines in GDP growth, then a suitable weight function to employ in the weighted multivariate scores is $w(\mathbf{z}) = \ind\{ z_{1} < 0, z_{2} < 0 \}$. This weight function can readily be extended to consider further quarters of negative GDP growth. 

The threshold-weighted energy score, variogram score, and maximum mean discrepancy score are all shown below for the Markov switching autoregressive forecasts when predicting negative GDP growth in the two following quarters. 

\begin{Schunk}
\begin{Sinput}
R> d <- 2
R> scores_twes <- twes_sample(obs[1:d], X[1:d, ], b = 0)
R> scores_twvs <- twvs_sample(obs[1:d], X[1:d, ], b = 0)
R> scores_twmmds <- twmmds_sample(obs[1:d], X[1:d, ], b = 0)
R> print(cbind(scores_twes, scores_twvs, scores_twmmds))
\end{Sinput}
\begin{Soutput}
     scores_twes scores_twvs scores_twmmds
[1,]        1.77        2.62         0.242
\end{Soutput}
\end{Schunk}

Note that these weighted scoring rules consider not only the probability that a recession will occur, but also the severity of the recession.

\section{Summary and discussion}
\label{sec:summary}

Scoring rules are well-established when evaluating and comparing probabilistic forecasts, and the \pkg{scoringRules} package in \proglang{R} has become well-established when implementing popular scoring rules in practice. In this paper, we discuss how the functionality of the \pkg{scoringRules} package has been extended such that particular outcomes can be emphasised when using scoring rules to assess forecast performance. 

Two approaches to target particular outcomes are available, which can be applied to probabilistic forecasts for both univariate and multivariate outcomes. These approaches are available for popular scoring rules including the LogS, CRPS, ES, VS, and MMDS, facilitating very flexible, user-oriented evaluation of probabilistic forecasts in a wide range of practical applications. In particular, functionality is available to calculate the conditional and censored likelihood scores proposed by \cite{DiksEtAl2011}; outcome-weighted versions of the CRPS, ES, VS, and MMDS, which can be constructed from the general framework outlined by \cite{HolzmannKlar2017}; and threshold-weighted versions of the CRPS, ES, VS, and MMDS \citep{MathesonWinkler1976, GneitingRanjan2011, AllenEtAl2022}.

While the \pkg{scoringRules} package contains analytical expressions for the LogS and CRPS for several parametric distributions, the weighted scoring rules discussed herein are only available for forecast distributions in the form of a simulated sample, or an ensemble. In this case, the weighted scoring rules can readily be calculated for arbitrary weight functions, which is generally not the case for forecasts in the form of parametric distributions. While this permits very flexible forecast evaluation, the \pkg{scoringRules} package could be extended further by incorporating weighted scoring rules for certain families of parametric distributions.

\section*{Acknowledgements}

This work was funded by the Swiss Federal Office for Meteorology and Climatology (MeteoSwiss) and the Oeschger Centre for Climate Change Research. I am very grateful to Fabian Kr{\"u}ger, Sebastian Lerch, and Alexander Jordan for their contribution to this work. Jonas Bhend and Jos{\'e} Carlos Araujo Acu{\~n}a are also thanked for their many fruitful suggestions.

\bibliography{bibliography_weighted}

\begin{appendix}
	\clearpage
	\section{Scores for simulated predictive distributions}\label{app:sampleformulas}
	
	Consider $\Omega \subseteq \R$, and suppose the forecast distribution is only available via a simulated sample $x_{1}, \dots, x_{m} \in \Omega$. To evaluate the empirical distribution function defined by this sample, 
	\begin{equation*}
	  \hat{F}_{m}(z) = \frac{1}{m} \sum_{i=1}^{m} \ind \{x_{i} \leq z\},
	\end{equation*}
the CRPS simplifies to 
  \begin{equation*}
	  \mathrm{CRPS}(\hat{F}_{m}, y) = \frac{1}{m} \sum_{i=1}^{m} |x_{i} - y| - \frac{1}{2 m^{2}} \sum_{i=1}^{m} \sum_{j=1}^{m} |x_{i} - x_{j}|.
	\end{equation*}

  The outcome-weighted CRPS and threshold-weighted CRPS are defined analogously: given a univariate weight function $w$, the outcome-weighted CRPS can be written as
 \begin{equation*}
	  \mathrm{owCRPS}(\hat{F}_{m}, y) = \frac{1}{m \bar{w}} \sum_{i=1}^{m} |x_{i} - y| w(x_{i}) w(y) - \frac{1}{2 m^{2} \bar{w}^{2}} \sum_{i=1}^{m} \sum_{j=1}^{m} |x_{i} - x_{j}| w(x_{i}) w(x_{j}) w(y),
	\end{equation*}
  where $\bar{w} = \sum_{i=1}^{m} w(x_{i})/m$. Letting $v$ denote the corresponding chaining function, the threshold-weighted CRPS is
\begin{equation*}
	  \mathrm{twCRPS}(\hat{F}_{m}, y) = \frac{1}{m} \sum_{i=1}^{m} |v(x_{i}) - v(y)| - \frac{1}{2 m^{2}} \sum_{i=1}^{m} \sum_{j=1}^{m} |v(x_{i}) - v(x_{j})|.
	\end{equation*}

  Similarly, let $F$ be a forecast distribution on $\Omega \subseteq \R^{d}$ for $d > 1$, and suppose that only a sample $\bx_{1}, \dots, \bx_{m}$ from $F$ is available, with $\bx_{i} = \left( x_{i,1}, \dots, x_{i, d} \right) \in \Omega$ for $i = 1, \dots, m$. In this case, the energy score for the corresponding empirical multivariate distribution $\hat{F}_{m}$ can be written as 
  \begin{equation*}
    \mathrm{ES}(\hat{F}_{m}, \by) = \frac{1}{m} \sum_{i=1}^{m} \| \bx_{i} - \by \| - \frac{1}{2 m^{2}} \sum_{i=1}^{m} \sum_{j=1}^{m} \| \bx_{i} - \bx_{j} \|,
  \end{equation*}
  the variogram score of order $p$ becomes
  \begin{equation*}
    \mathrm{VS}^{p}(\hat{F}_{m}, \by) = \sum_{i=1}^{d} \sum_{j=1}^{d} h_{i,j} \left( \frac{1}{m} \sum_{k=1}^{m} \left|x_{k, i} - x_{k, j} \right|^{p} - \left|y_{i} - y_{j} \right|^{p} \right)^2,
  \end{equation*}
  and the maximum mean discrepancy score is 
  \begin{equation*}
    \mathrm{MMDS}(\hat{F}_{m}, \by) = \frac{1}{2 m^{2}} \sum_{i=1}^{m} \sum_{j=1}^{m} \mathrm{exp} \left\{ -\frac{1}{2} \| \bx_{i} - \bx_{j} \|^{2} \right\} - \frac{1}{m} \sum_{i=1}^{m} \mathrm{exp} \left\{ -\frac{1}{2} \| \bx_{i} - \by \|^{2} \right\}.
  \end{equation*}
\end{appendix}

 Given a multivariate weight function $w$ and chaining function $v$, the outcome-weighted and threshold-weighted versions of these scores can be calculated as follows. In this case, $\bar{w} = \sum_{i=1}^{m} w(\bx_{i})/m$, and $v(\bx_{i}) = \left( v(\bx_{i})_{1}, \dots, v(\bx_{i})_{d} \right) \in \Omega$ for $i = 1, \dots, m$.
 \begin{equation*}
    \mathrm{owES}(\hat{F}_{m}, \by) = \frac{1}{m \bar{w}} \sum_{i=1}^{m} \| \bx_{i} - \by \| w(\bx_{i}) w(\by) - \frac{1}{2 m^{2} \bar{w}^{2}} \sum_{i=1}^{m} \sum_{j=1}^{m} \| \bx_{i} - \bx_{j} \| w(\bx_{i}) w(\bx_{j}) w(\by);
  \end{equation*}
 \begin{equation*}
    \mathrm{owVS}^{p}(\hat{F}_{m}, \by) = w(\by) \sum_{i=1}^{d} \sum_{j=1}^{d} h_{i,j} \left( \frac{1}{m \bar{w}} \sum_{k=1}^{m} \left|x_{k,i} - x_{k,j} \right|^{p} w(\bx_{k}) - \left|y_{i} - y_{j} \right|^{p} \right)^2;
  \end{equation*}
 \begin{equation*}
 \begin{split}
    \mathrm{owMMDS}(\hat{F}_{m}, \by) =& \frac{1}{2 m^{2} \bar{w}^{2}} \sum_{i=1}^{m} \sum_{j=1}^{m} \mathrm{exp} \left\{ -\frac{1}{2} \| \bx_{i} - \bx_{j} \|^{2} w(\bx_{i}) w(\bx_{j}) w(\by) \right\} \\
    &- \frac{1}{m \bar{w}} \sum_{i=1}^{m} \mathrm{exp} \left\{ -\frac{1}{2} \| \bx_{i} - \by \|^{2} w(\bx_{i}) w(\by) \right\};
  \end{split}
  \end{equation*}
 \begin{equation*}
    \mathrm{twES}(\hat{F}_{m}, \by) = \frac{1}{m} \sum_{i=1}^{m} \| v(\bx_{i}) - v(\by) \| - \frac{1}{2 m^{2}} \sum_{i=1}^{m} \sum_{j=1}^{m} \| v(\bx_{i}) - v(\bx_{j}) \|;
  \end{equation*}
	\begin{equation*}
    \mathrm{twVS}^{p}(\hat{F}_{m}, \by) = \sum_{i=1}^{d} \sum_{j=1}^{d} h_{i,j} \left( \frac{1}{m} \sum_{k=1}^{m} \left|v(\bx_{k})_{i} - v(\bx_{k})_{j} \right|^{p} - \left|v(\by)_{i} - v(\by)_{j} \right|^{p} \right)^2;
  \end{equation*}
  \begin{equation*}
    \mathrm{twMMDS}(\hat{F}_{m}, \by) = \frac{1}{2 m^{2}} \sum_{i=1}^{m} \sum_{j=1}^{m} \mathrm{exp} \left\{ -\frac{1}{2} \| v(\bx_{i}) - v(\bx_{j}) \|^{2} \right\} - \frac{1}{m} \sum_{i=1}^{m} \mathrm{exp} \left\{ -\frac{1}{2} \| v(\bx_{i}) - v(\by) \|^{2} \right\}.
  \end{equation*}
\end{document}